\documentclass[aps,pre,onecolumn,superscriptaddress]{revtex4}

\pdfoutput=1
\usepackage{graphicx}
\usepackage{enumerate}

\newcommand{\dx}{h}

\begin{document}
\title{Properties of pedestrians walking in line -- Fundamental diagrams}
\author{Asja Jeli\'c}
\email[]{asja.jelic@gmail.com}
\affiliation{Laboratory of Theoretical Physics,
CNRS (UMR 8627), University Paris-Sud\\
Batiment 210, F-91405 Orsay Cedex, France.}
\author{C\'ecile Appert-Rolland}
\email[]{cecile.appert-rolland@th.u-psud.fr}
\affiliation{Laboratory of Theoretical Physics,
CNRS (UMR 8627), University Paris-Sud\\
Batiment 210, F-91405 Orsay Cedex, France.}
%\thanks{author to whom correspondence should be addressed}
\author{Samuel Lemercier}
\email[]{samuel.lemercier@inria.fr}
\affiliation{INRIA Rennes - Bretagne Atlantique, Campus de Beaulieu\\
35042 Rennes, France.}
\author{Julien Pettr\'e}
\email[]{julien.pettre@inria.fr}
\affiliation{INRIA Rennes - Bretagne Atlantique, Campus de Beaulieu\\
35042 Rennes, France.}

\date{\today}

\begin{abstract}
We present experimental results obtained for a one-dimensional
pedestrian flow using high precision motion capture.
The full pedestrians’ trajectories are obtained.
In this paper, we focus on the fundamental diagram,
and on the relation between the instantaneous velocity and
spatial headway (distance to the predecessor).
While the latter was found to be linear in previous experiments,
we show that it is rather a piecewise linear behavior which is
found if larger density ranges are covered. 
Indeed, our data clearly exhibits three
distinct regimes in the behavior of pedestrians
that follow each other.
The transitions between these regimes
occur at spatial headways
of about $1.1$ and $3$ m, respectively.
This finding could be useful for future modeling.
\end{abstract}

\pacs{}

\maketitle

%\newpage

\section{Introduction}

There is a growing interest to understand
the behavior of pedestrians, not only
in emergency regimes - which can represent
a real danger for the persons involved - but
also in normal pedestrians flows, for example
to improve the capacity of a pedestrian facility.
But there is also a more fundamental interest
for physicists to study pedestrian crowds,
as these exhibit
collective phenomena and show 
several types of pattern formation.
Several models were proposed to describe
the dynamics of pedestrians \cite{helbing_m95,burstedde01b}
which partially
reproduce the behavior of real crowds.
However, several intricate effects have to be
taken into account, in particular due to the
bidimensional nature of pedestrian traffic
(compared to the quasi one-dimensional road traffic, for
example).
Pedestrians avoid each other by adapting
both their velocity amplitude and direction
\cite{paris_p_d07,ondrej10b}.
It is tempting to separate these effects
by considering some simplified configurations.

In particular, one-dimensional pedestrian flows
allow one to restrict oneself to longitudinal interactions,
and to avoid lateral ones. Such longitudinal
interactions may occur in narrow corridors
\cite{chraibi_s_s10},
where pedestrians cannot pass each other,
or in larger corridors when the encounter
of oppositely moving pedestrians induces the
formation of several (possibly narrow) lanes
\cite{moussaid11,zhang12}.
The interpretation of two-dimensional data
may require an understanding of the following
behavior \cite{johansson09}.
One-dimensional flows can be used to test
the following behavior in models \cite{seyfried_s_l06}.
Besides, some models do use a decoupling between the
angular and speed adaptation
\cite{ondrej10b,moussaid_h_t11a,lemercier12a}.

If the line on which pedestrians walk is closed,
boundary effects are avoided and it becomes
easier to study the bulk behavior.
Such experiments have indeed already been performed
in the past.
In \cite{jezbera10}, the distribution
of time headways was measured.
Seyfried {\em et al.} \cite{seyfried05} have measured
the density and velocity of pedestrians 
following an oval trajectory.
They considered densities up to 2 ped/m,
where ped means pedestrians, and
were able to draw the corresponding fundamental diagram \footnote{A fundamental
diagram refers to a plot of either the velocity or the flux
of some moving agents (pedestrians, cars, ants....)
as a function of density.}.
This experiment was compared to a similar one
performed in India \cite{chattaraj_s_c09}.
In both cases, data were obtained
from localized measurements.

In this paper, we present new experimental results
that were obtained on a circular path using high precision 
motion capture \footnote{This experiment belongs to a series
of experiments that was realized in the frame of the French {\sc PEDIGREE} Project \cite{pedigree_info}.}.
The trajectories 
of all pedestrians 
were reconstructed in three-dimensions
for the whole duration of the experiments.
As a result, we were able to obtain not only measurements
that can be directly compared to previous studies,
but we had access to new features.

In particular, we can now obtain an {\em instantaneous}
fundamental diagram, giving the relation between
the instantaneous velocity of a pedestrian and
his instantaneous density, defined as the inverse
of his spatial headway $\dx$.
We show that such a fundamental diagram is a mixture
between stationary behavior,
and transient
behavior, for which the velocity does not have time
to relax toward its stationary value.
This may occur for example when a pedestrian arrives
on the rear end of a jam, or on the contrary when
it leaves the jam. 

Our main result is that the behavior of pedestrians in
a following regime
exhibits two transitions, 
one occurring at a spatial headway
of about $1.1$ m and again another transition around $3$ m.
These transitions are clearly visible on the velocity-spatial headway relation $v(\dx)$.
Actually, this relation was found to be
linear in previous experiments \cite{seyfried05,chattaraj_s_c09}.
Our experiment has allowed us to access a much larger 
density range, and revealed that actually it is a piecewise
linear relation.

We shall first describe the experiment in section \ref{sect_exp}.
Experimental measurements of the fundamental diagram
will be presented in section \ref{sect_fd}. We shall discuss
in detail global, local (subsection \ref{sect_global}), and instantaneous
fundamental diagrams (subsection \ref{sect_instantaneous}).
In section \ref{sect_transition}, we shall present our main result, showing
that there exists two transitions in the behavior of pedestrians
that can be seen in the 
velocity-spatial headway relation $v(\dx)$.
Our results will then be compared with fundamental diagrams
from other countries, in order to discuss possible cultural variations
(section \ref{sect_culture}).
We conclude in section \ref{sect_conclusion}.

\section{The experiment}
\label{sect_exp}

The goal of our experiment was to characterize
one-dimensional pedestrian traffic at
various densities.
Different densities were obtained by modifying
on the one hand the number of pedestrians involved,
and on the other hand the path on which they walk.

\begin{figure}[t] %  figure placement: here, top, bottom, or page
  \centering
   \includegraphics[width=0.8\textwidth]{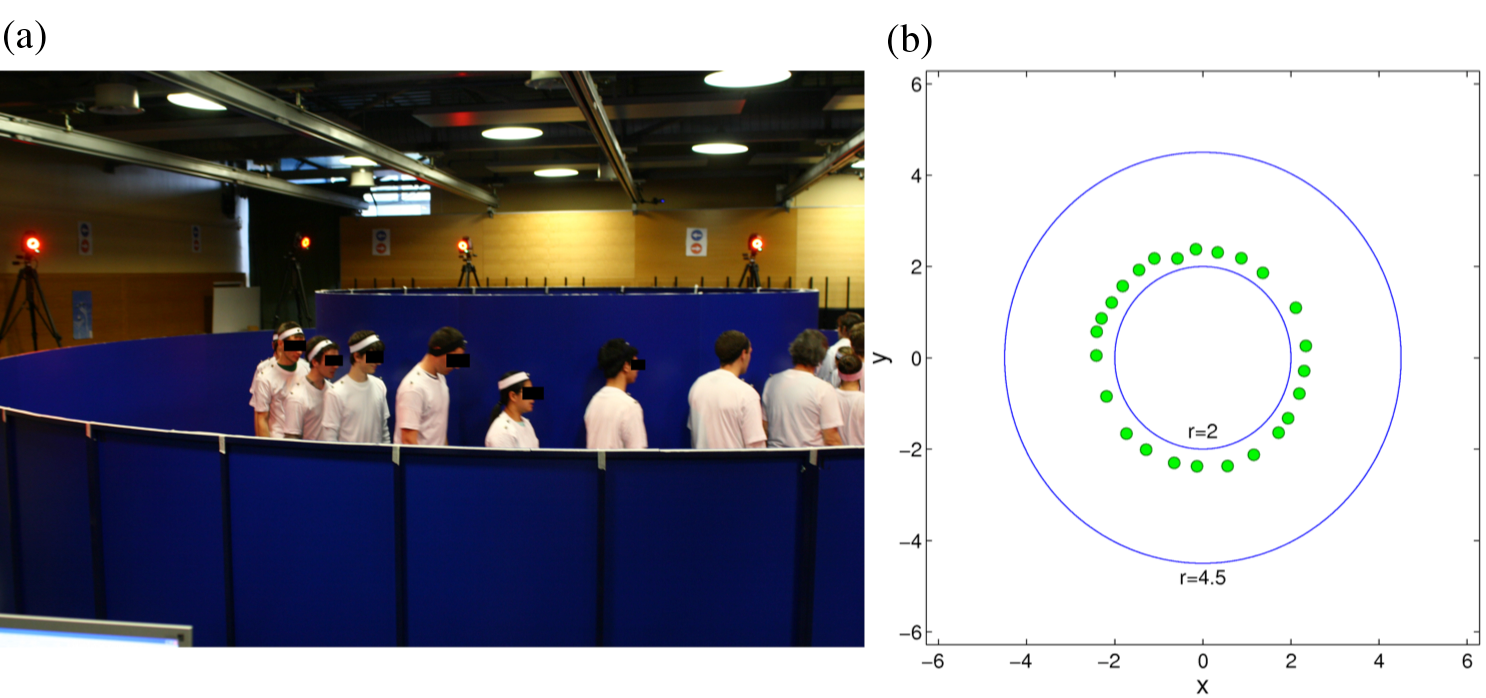}
   \caption{
   (Color online)
   Experimental set-up: (a) photograph of the experiment and (b)
   top view obtained through {\sc MATLAB} treatment of the data,
   showing a snapshot of the pedestrians positions (small circles).
   Both images show an experiment along the inner circle.
 }
   \label{fig:ring}
\end{figure}

Up to 28 pedestrians (20 males and 8 females)
were involved in the experiment.
The experiment was conducted inside a ring corridor
formed by inner and outer circular walls of radii
$2$ and $4.5$ m, respectively (see figure \ref{fig:ring}).
Participants
were told to walk in line along either the inner or outer
wall, without passing each other. 
This resulted into two types of pedestrians trajectories:
the inner circular path of observed average radius $2.4$ m and
length $15.08$ m, and the outer circular path of observed
average radius $4.1$ m and length $25.76$ m.
Pedestrians were volunteers, 
and all were naive about the purpose of the experiment. 
They were asked to walk in a ``natural way'',
as if they were walking alone in the street (and without
talking to each other).

As a result of the variation of the number of
pedestrians involved and of the use of the inner or outer
circular path, the average density of pedestrians
varied from $0.31$ to $1.86$ ped/m.

Table \ref{table:tab_nb} gives a summary of the parameter
values for which experiments were performed, 
and indicates the number of replicas
that were performed for each parameter set. 
When it was possible (i.e.,\ for low densities), 
replicas were involving different pedestrian sets.
For half of the replicas, pedestrians were walking
clockwise, and counterclockwise for the other half.
The typical duration of an experiment was $1$ minute
and somewhat more for higher densities.

\begin{table}
\begin{tabular}{|c|c|c|c|}
\hline
number of & circle  & density & number of \\[-0.0cm]
pedestrians & & [ped/m] & replicas \\
\hline
8  & outer & 0.31 & 8\\
16 & outer & 0.62 & 6\\
24 & outer & 0.93 & 4\\
16 & inner & 1.06 & 8\\
28 & outer & 1.09 & 5\\
20 & inner & 1.32 & 5\\
21 & inner & 1.39 & 4\\
24 & inner & 1.59 & 8\\
28 & inner & 1.86 & 4\\
\hline
\end{tabular}
\caption{Number of experimental replicas for each
set of parameters (sorted by density):
number of pedestrians, 
use of the inner or outer circular path,
and density of pedestrians.}
\label{table:tab_nb}
\end{table}

The whole experiment was tracked by a VICON MX-$40$ motion capture system.
Each pedestrian was equipped with $4$ markers,
one on the left shoulder, two on the right shoulder,
and one on top of head.
The experimental setting was surrounded by 12 infra-red
cameras able to detect the markers.
The reconstruction software {\sc VICON IQ} was used to convert
the raw data into three-dimensional (3D) trajectories for each of the markers, with a frequency of 
120 frames/s. The reconstruction is described
in more details in \cite{lemercier11b}.
From the position of the four markers, a mean position
point is calculated for each pedestrian (we shall refer
to this point as the 'center of mass' of the pedestrian
in the following,
though it is more a mean position).
Besides, in the analysis presented in this paper, we filtered out
the step oscillations (steps characteristics will be studied
in a next paper \cite{jelic11b}).

In the data analysis, we had to take into account
that some markers may be occluded at some times.
Indeed it may happen that, due to the experimental setup, some markers
are hidden by the walls or pedestrians' bodies. 
In particular for high densities, the markers on the
shoulder closest to the wall were most of the time
not visible.
In the following analysis, we carefully checked to keep 
only data for which we had enough confidence.

\section{Fundamental diagram}
\label{sect_fd}

The fundamental diagram, which gives flow or velocity
as a function of density, is the main quantity
that is usually measured in road or pedestrian systems.
Indeed, it immediately shows transitions from free
flow to jammed states.
Besides, it allows for an easy comparison between
different systems.
It is also often used as a basic ingredient for modeling.
But in fact, for pedestrians, the fundamental diagram was shown
to vary significantly depending on the precise geometry within which
pedestrians were walking \cite{zhang11,daamen_h03b}.

Thus we shall compare our findings only with one-dimensional
experiments, though it seems that comparisons with two-dimensional
flows could be possible \cite{seyfried05}.
For previous one-dimensional experiments \cite{seyfried05},
measurements were performed
locally on a window of the trajectory, and each data point
was an average over this measurement window.
With our data, at least three different definitions can
be chosen for the density and velocity, depending on the
way averages are performed, and ranging from global
to instantaneous measurements.

\begin{enumerate}[(i)]
\item
Global measurements: fluxes, velocities and densities are
averaged over the whole system and the whole duration 
of experiment (excluding the initial and final transients).
Densities are thus discrete and correspond to the few 
densities imposed in the experiments.
Both the average flux or average velocity can be plotted as a function 
of density.

\item
Local measurements: velocities and densities are
measured in a subpart of the system, and averaged 
over a finite time period.
Local densities may vary around the global value.
It is this type of measurements that can be compared to the 
experiments by Seyfried {\em et al.} \cite{seyfried05,seyfried07b}.
As the way averages are performed may affect the shape
of the fundamental diagram \cite{schadschneider_s11},
we followed exactly the same procedure as in \cite{seyfried05}
to determine the local density and velocity, in order to
allow for a direct comparison among experiments. 

\item
Instantaneous individual measurements: if the distance between
two pedestrians is interpreted as the inverse of
a density, then the instantaneous velocity of
a given pedestrian can be plotted as a function
of the inverse distance that is available in
front of him
\footnote{We neglect the thickness of people 
in these distance measurements, i.e. the distance
that we consider is actually the distance between
the centers of mass of successive pedestrians. Indeed
we do not have any information about the real size
of each pedestrian.}.
This gives another type of fundamental diagram,
which spans over a much larger range of densities,
from 0.1 up to about 3 ped/m, as we shall see later.
\end{enumerate}

\subsection{Global and local fundamental diagrams}
\label{sect_global}

\begin{figure}[t] %  figure placement: here, top, bottom, or page
  \centering
   \includegraphics[width=0.6\textwidth]{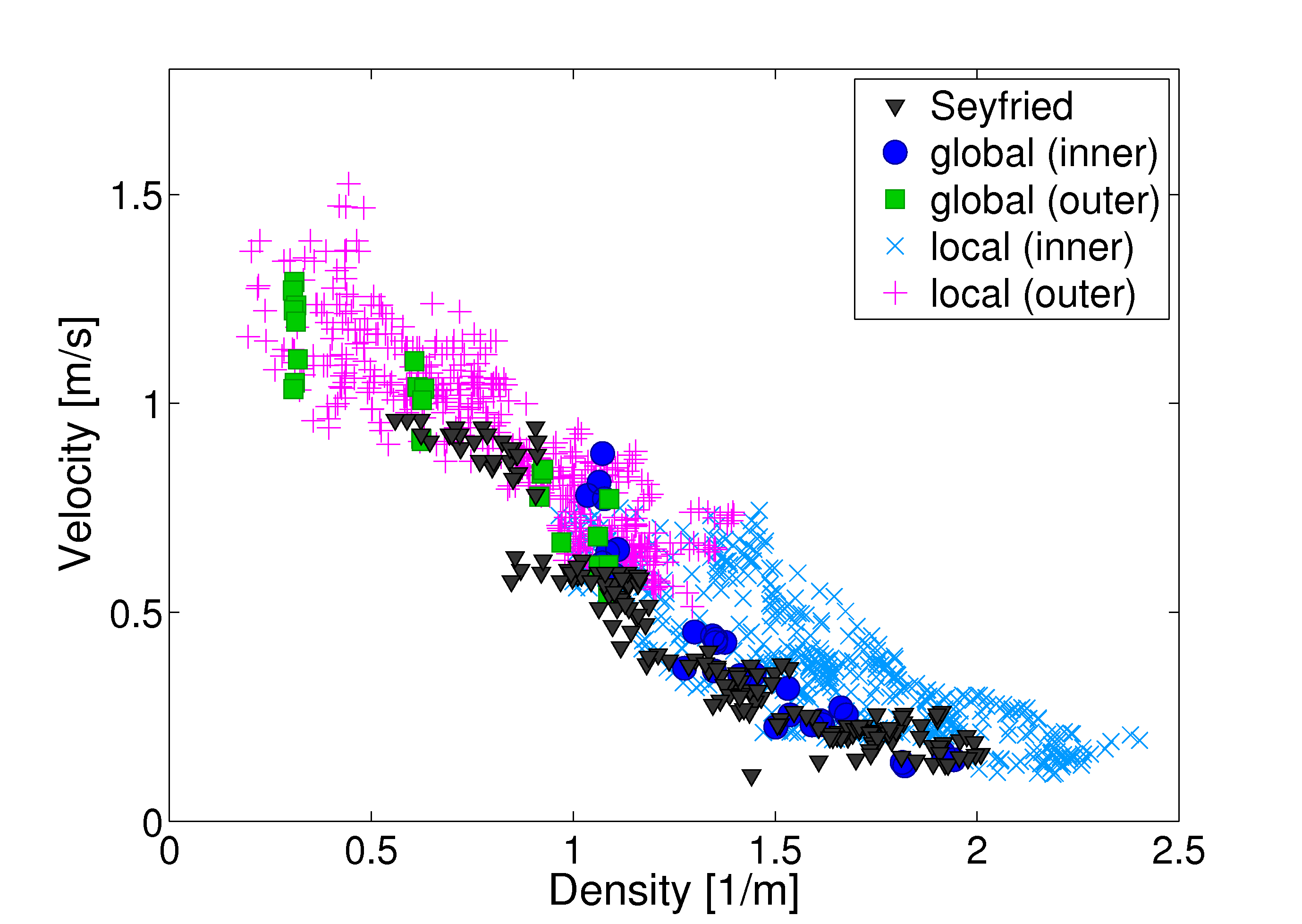}\\
   \caption{
   (Color online)
Fundamental diagram: Global and local measurements,
compared with the measurements by Seyfried {\em et al.}
\cite{seyfried05,seyfried07b};
Data were obtained either along the inner circular
path (circle and ‘x’ symbols) or along the outer circular path (square
and ‘+’ symbols).
 }
   \label{fig:fd_global1}
\end{figure}

In this subsection,  
we present the results of the 
global and local measurements of the 
velocity as a function of density
(figure \ref{fig:fd_global1}),
and compare them to the experiments by Seyfried 
{\em et al.}\ \cite{seyfried05,seyfried07b}.

\vskip 0.5cm
{\bf Measurement procedures}\nopagebreak
\vskip 0.5cm

Note that in the case of global measurements, experiments
with the same number of pedestrians do not necessarily
correspond exactly to the same density.
Indeed, global densities are calculated
for each experiment by dividing a given number of participants $N$
with the length of the circular path obtained using the measured average
radius of all pedestrians during the whole duration $t$ of the experiment.
We denote this average as $\left< \right>_{N,t}$, so that 
\begin{equation}
\rho = \frac{N}{2\pi \left< R\right>_{N,t}}.
\end{equation}
This procedure gives a small dispersion in the global values 
of density for experiments with the same number of pedestrians.

Measurement of the local velocity and density is performed
by following the same procedure as in \cite{seyfried05}.
The local mean velocity and density are calculated by considering 
a segment $\Delta \theta$ of the circle as a measurement area.
The velocity of each pedestrian $j$ passing through this area is defined
as the length $x=\Delta \theta \left<R\right>_{N, t}$ divided by 
the time he or she needs to cross this circle area 
\begin{equation}
v_{j} = \frac{x}{t^{o}_{j}-t^{i}_{j}},
\end{equation}
where $t^{i}_{j}$ and $t^{o}_{j}$ are the in and out times at which a pedestrian $j$ 
enters and exits the measurement area, respectively.
Density of pedestrians inside the measurement area at time $t$
is given by $\rho(t)=\sum_{j}\Theta_{j}(t)/x$, where $\Theta_{j}(t)$
is the fraction of space between pedestrian $j$ and his follower 
$j+1$ that falls inside the measurement area
\begin{equation}
\Theta_{j}(t) = \left\{ 
 \begin{array}{ccc}
 \frac{t-t_{j}^{i}}{t_{j+1}^{i}-t_{j}^{i}}& \text{for} & t\in \left[t_{j}^{i},t_{j+1}^{i}\right]  \\
1 & \text{for} & t\in \left[t_{j+1}^{i},t_{j}^{o}\right] \\
  \frac{t_{j+1}^{o}-t}{t_{j+1}^{o}-t_{j}^{o}}& \text{for} & t\in \left[t_{j}^{o},t_{j+1}^{o}\right] \\
0 &  & \text{otherwise.}
 \end{array}\right. 
\end{equation}
The density $\rho_{j}$ for each person $j$ is then calculated as 
$\rho_{j}=\left<\rho(t)\right>_{{\Delta t}_{j}}$, i.e.,\
as an average of the density in the measured area over the time interval 
$\Delta t_{j}=t^{o}_{j}-t^{i}_{j}$.
Finally, in figure \ref{fig:fd_global1}, for each density $\rho_{j}$,
we plot the velocity $v_{j}$.

\vskip 0.5cm
{\bf Discussion}\nopagebreak
\vskip 0.5cm

\begin{figure}[t] %  figure placement: here, top, bottom, or page
  \centering
   \includegraphics[width=0.6\textwidth]{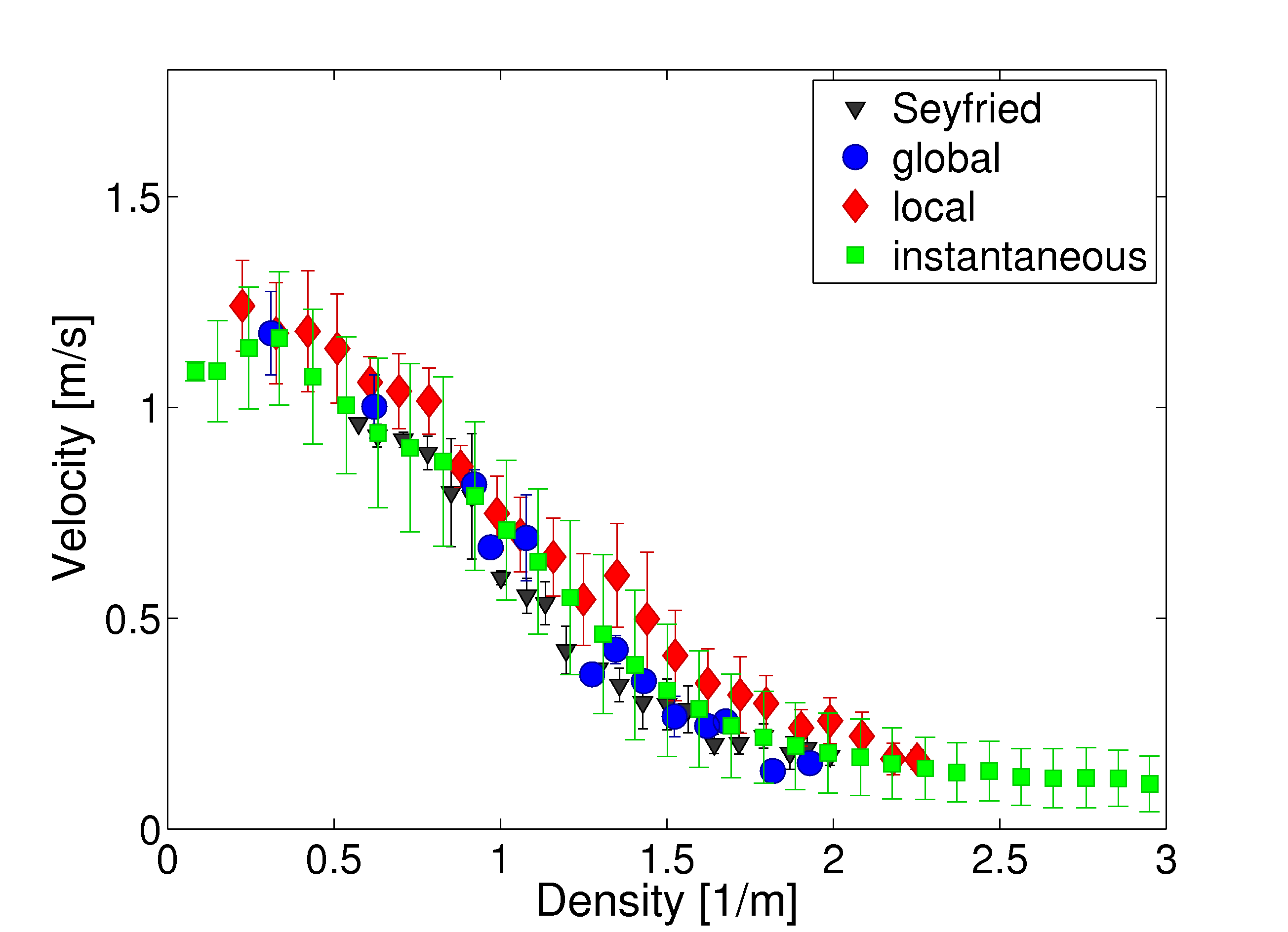}
   \caption{
   (Color online)
Fundamental diagram: 
results of the binning procedure applied to 
the global, local, and instantaneous measurements,
as well as to 
the measurements by Seyfried {\em et al.}
Bars give the standard deviation.
 }
   \label{fig:fd_global2}
\end{figure}

Let us first consider the global fundamental diagram
shown in figure \ref{fig:fd_global1}.
For similar densities, one notices
a dispersion of the global velocity values among different replicas.
The dispersion is stronger for the lower densities.
We explain this by the fact that at low densities, different
individuals were involved in the various experimental replicas,
therefore possibly imposing different preferred velocities.
For higher densities,
the same pedestrians could be part of
several experimental replicas. Besides
global constraints restrict more the velocity choice
of each pedestrian when density increases.
As a result, dispersion is much lower for high densities.
We see the same dispersion reduction on local data
in figure \ref{fig:fd_global2}, once a binning procedure
has been applied.

Figure \ref{fig:fd_global1} shows that, as expected, the velocity 
decreases with increasing density.
Actually, when the density increases, some stop-and-go waves
are formed.
It is possible with our experimental set-up to detect
these waves very precisely
\cite{lemercier11a,lemercier12a}.
A thorough study of such stop-and-go waves
will be presented later.

In order to compare the fundamental diagram more easily
than through data clouds, we have applied a binning procedure
to the data, with the result shown in figure \ref{fig:fd_global2}.
Our results are in good qualitative agreement with Seyfried's \cite{seyfried05},
though the velocities  that we find are systematically
slightly above.

We find that our global and local measurements give quite close
results as long as the density is smaller than 1.2 ped/m,
while both measurements differ for larger
densities, i.e. when stop-and-go waves arise.

\subsection{Instantaneous fundamental diagram}
\label{sect_instantaneous}

As we have the full trajectories of all pedestrians at all times,
we can also measure some 'instantaneous' fundamental diagram.

\vskip 0.5cm
{\bf Measurement procedure}\nopagebreak
\vskip 0.5cm

More precisely, for each pedestrian, 
at each time frame, we measure the instantaneous velocity,
and the distance $\dx$ between the 
centers of mass of the pedestrian under consideration and
his predecessor. We call this distance a 'spatial headway',
though it neglects the size of pedestrians. The inverse of
this distance is interpreted as an instantaneous
density.

Then, for a given pedestrian, 
these instantaneous density and velocity values are 
averaged over $60$ consecutive time frames (i.e. over
a time window of $0.5$ seconds),
generating one data point.
The final fundamental diagram is 
obtained from the set of all the data points
through a binning procedure.

\begin{figure}[t] %  figure placement: here, top, bottom, or page
  \centering
   \includegraphics[width=0.48\textwidth]{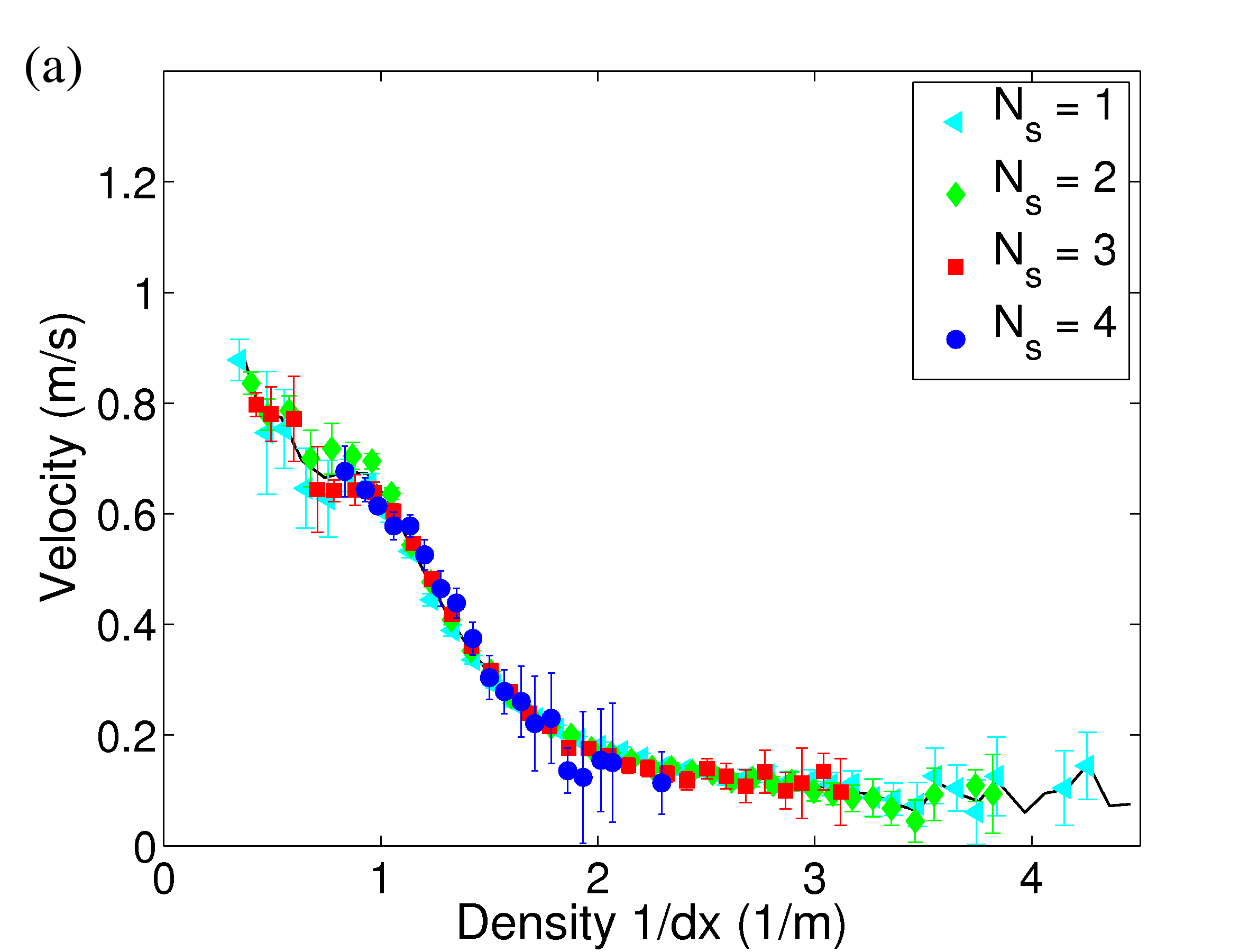}
   \includegraphics[width=0.48\textwidth]{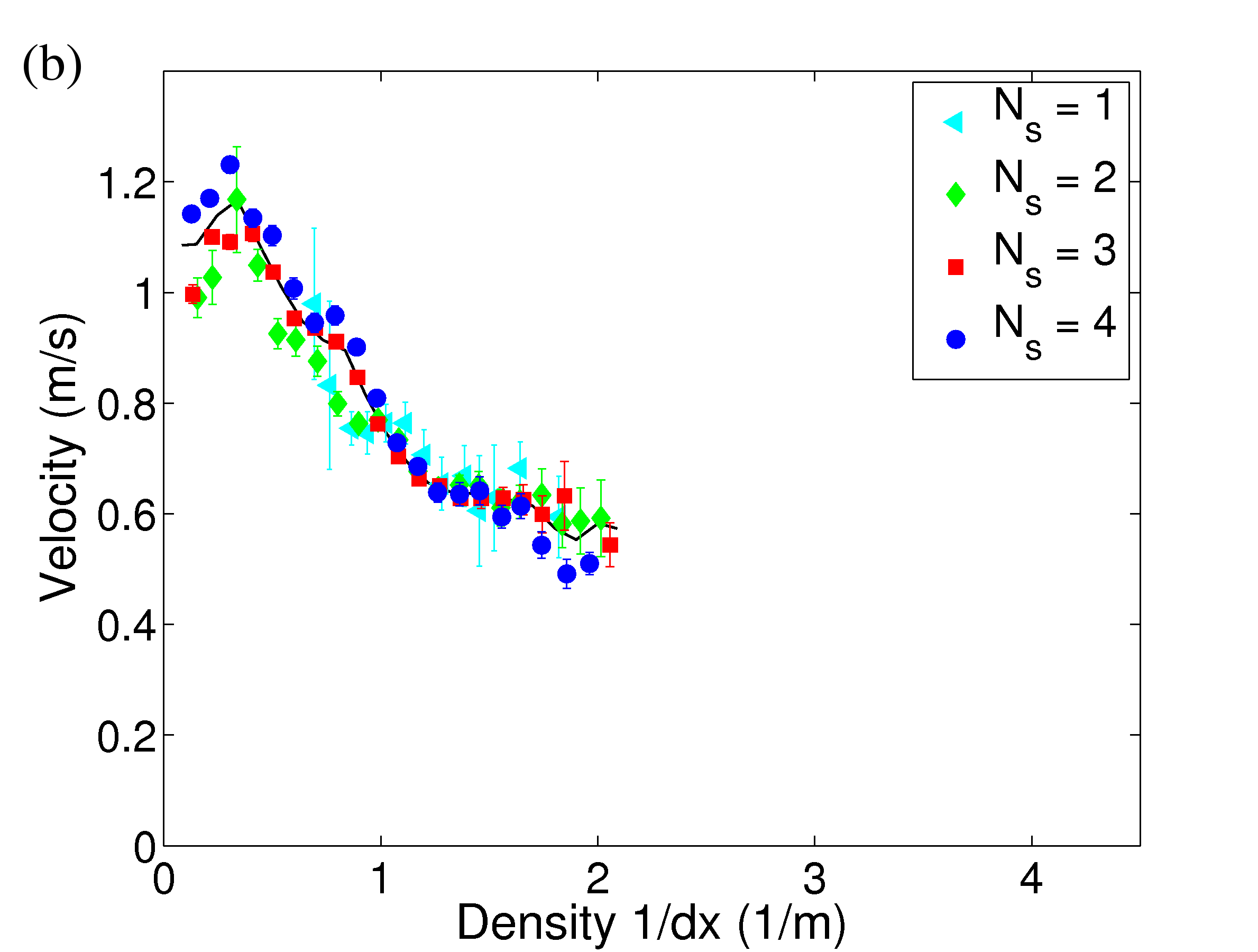}
   \caption{
   (Color online)
Fundamental diagram: Instantaneous velocity of individual pedestrians
as a function of density,
defined as the inverse of the instantaneous distance to the predecessor.
Data are sorted by the minimal number of detected markers per
pedestrians $N_s$ for the (a) inner and (b) outer
circular paths. We distinguish the cases 
$N_s=1$ (cyan triangles),
$N_s=2$ (green diamonds),
$N_s=3$ (red squares), and
$N_s=4$ (blue circles).
The solid line gives the average over all data, whatever the
number of markers is.
Error bars give the error on the mean.
 }
   \label{fig:fd_markers}
\end{figure}

\begin{figure}[t] %  figure placement: here, top, bottom, or page
  \centering
  \includegraphics[width=0.55\textwidth]{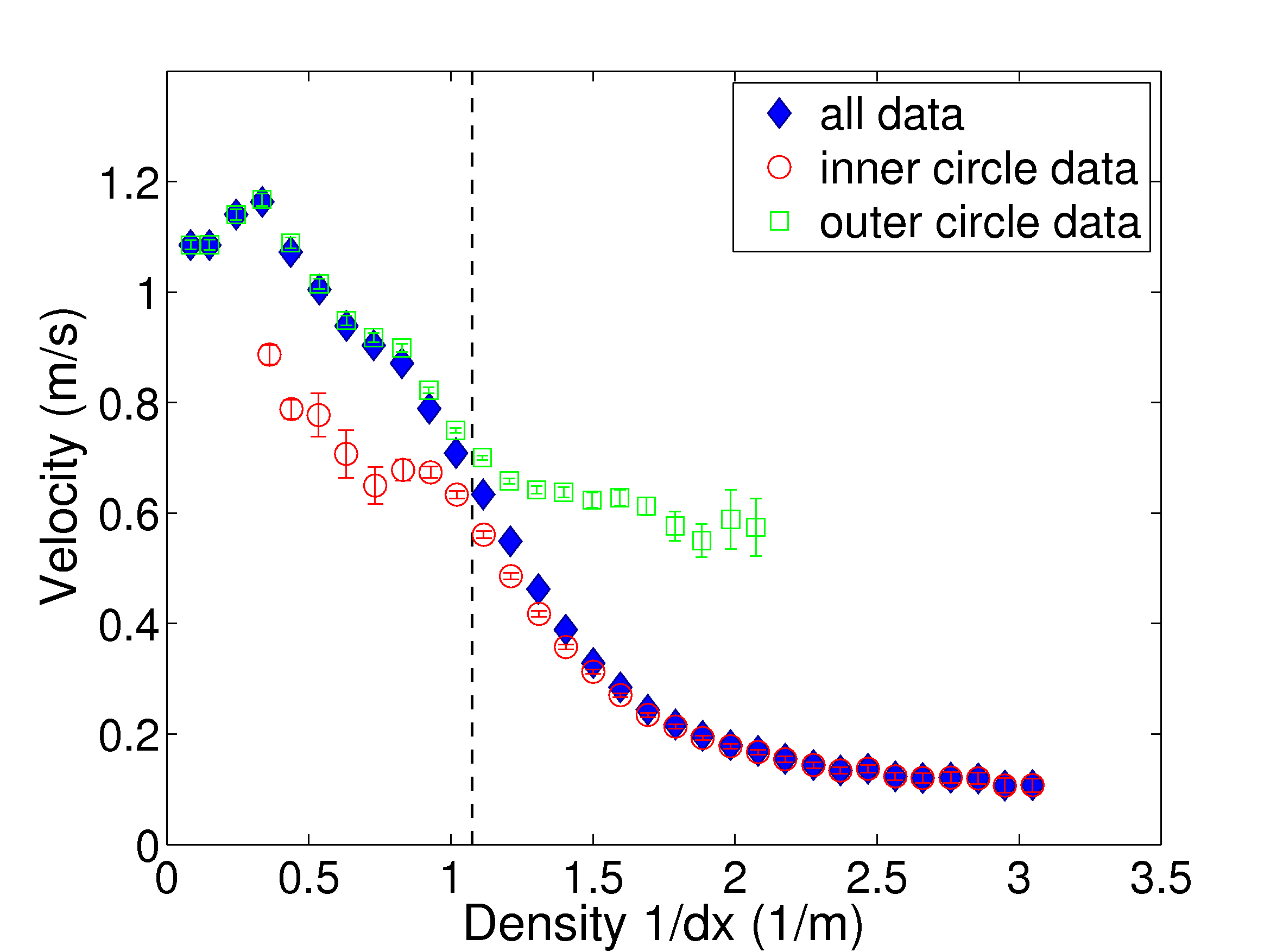}
   \caption{
   (Color online)
Fundamental diagram: Instantaneous velocity of individual pedestrians
as a function of density,
defined as the inverse of the instantaneous spatial headway
to the predecessor.
Data corresponding to the inner (outer) circle are plotted with
red circles (green squares), while the average over all data is
plotted with blue diamonds.
The vertical line indicates the maximal (minimal) global density
obtained on the outer (inner) circle (1.09 and 1.06 ped/m respectively).
Error bars give the error on the mean.
 }
   \label{fig:fd_markers_inout}
\end{figure}

Fundamental diagrams obtained from individual
instantaneous 
measurements are given in figures \ref{fig:fd_global2}, \ref{fig:fd_markers}
and \ref{fig:fd_markers_inout} and will be discussed below.

\vskip 0.5cm
{\bf Influence of the number of detected markers}\nopagebreak
\vskip 0.5cm

As discussed in section \ref{sect_exp}, at a given time frame
some or all markers of a given pedestrian may be occluded.
An interpolation is performed to generate data, but if
the time window on which markers are occluded is too long,
the interpolation procedure may introduce
some error in the interdistance measurements, which results
into unrealistic densities.

In the fundamental diagram of figure \ref{fig:fd_markers},
we distinguish experimental data depending on which number
of markers the measurement is based.
More precisely, we define $N_s$ as the 
minimum
number of markers that were detected,
at each of the consecutive $60$ time frames
over which we average, both for the observed pedestrian and his 
predecessor (with respect to which we determine the interdistance,
i.e.,\ inverse density).

We have plotted these data on different plots (figure \ref{fig:fd_markers}-a
or \ref{fig:fd_markers}-b) depending whether
they were obtained on the inner or outer circular path, respectively.

We find that the data obtained for different number of markers mostly overlap.
The main difference is that for low number of detected markers,
the measured density reaches higher values.
This seems to be an artefact of the measurements.
Indeed, when there are only one or two markers that are detected
for at least one of the pedestrians (predecessor or follower),
then it is not obvious in most cases to decide whether
these markers are on the head or on the shoulder.
As the head markers are, on average, 10 cm ahead compared
to shoulder markers, this creates an uncertainty on
the distance between pedestrians, which should be of
the order of 5 cm in the case of two detected markers (as at least one
of them is on the shoulder), and 10 cm
in the case of one marker. This uncertainty is negligible
at low densities, but becomes relevant at high densities.
In particular, it can explain why, on figure \ref{fig:fd_markers} (a),
data with one or two detected markers extend over densities
up to 4.2 and 3.7 ped/m, respectively, while the maximum
density measured with three detected markers is 3.1 ped/m:
the uncertainty on position corresponds exactly (quantitatively) to this 
maximum density difference.
As a consequence, we believe that the measurements beyond
a density of 3.1 ped/m cannot be trusted.

\begin{figure}[t] %  figure placement: here, top, bottom, or page
  \centering
   \includegraphics[width=0.55\textwidth]{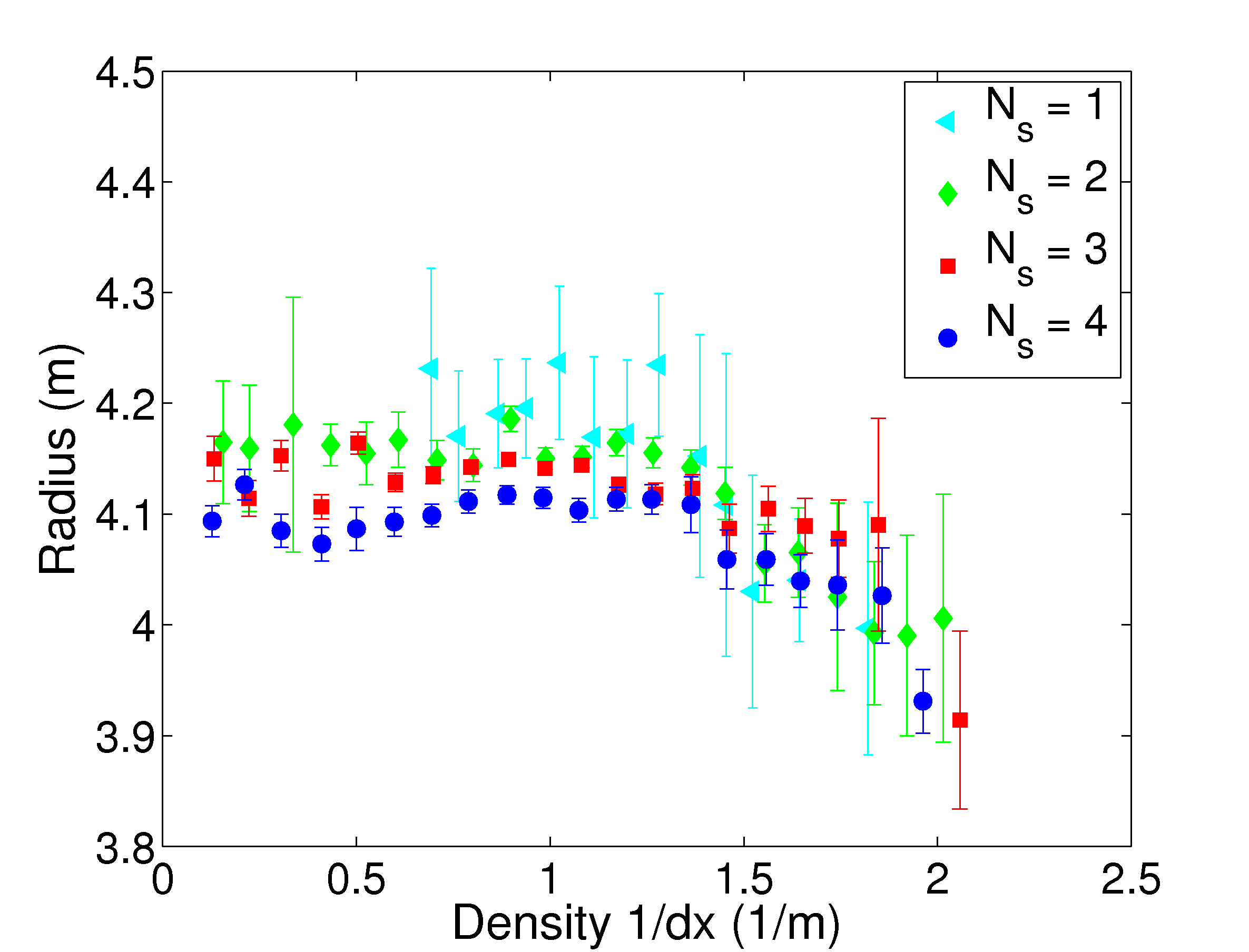}
   \caption{
   (Color online)
Radius as a function of the inverse spatial headway, for outer circle
experiments. The plot is obtained
as the result of a binning procedure over a large number of measurements.
Each measurement corresponds to an average over 0.5s, for one given pedestrian.
 }
   \label{fig:radius_markers}
\end{figure}

In most cases, only three markers are detected. 
This is due to the walls along which pedestrians are walking.
Figure \ref{fig:radius_markers} shows that on the outer
circle, pedestrians that were detected with their
four markers were on average walking at a larger distance
from the wall than others. Indeed it is more difficult to detect the
marker(s) on the wall's side when the pedestrian is very close
to the wall.
This is to be related to the fact that, 
for very low densities,
the velocities of pedestrians with four markers are larger
than those of pedestrians with three markers (see fig. \ref{fig:fd_markers}-b).

No such effect is observed on the inner circle, for which
there is no interest to escape from the wall, as it does not
shorten the trajectory. As a result, much less data with
four markers are available on the inner circle
(about 10 to 20 \% at high densities when pedestrians are
walking counterclockwise, and 5 to 10 \% when they are
walking clockwise, i.e. when the shoulder equipped with
two markers is close to the wall ).

\vskip 0.5cm
{\bf Discussion}\nopagebreak
\vskip 0.5cm

Figure \ref{fig:radius_markers} also shows that 
for higher densities along the outer circle, 
pedestrians tend to walk at a slightly larger
distance from the wall.
This can be an attempt to increase their visual horizon,
and also to limit the consequences of coming too close
to a collision with the predecessor.
Again, this effect is not present on the inner circle,
probably because pedestrians are more reluctant to escape
from the wall, as it lengthens their trajectory.

In figure \ref{fig:fd_markers_inout},
we superimpose the fundamental diagrams obtained for the inner
and outer circular paths.
The discrepancies that seem to arise do not actually indicate
significant differences in the behavior of pedestrians,
and we shall see below that they can be easily interpreted.
The average over all data (inner and outer circle)
goes smoothly over the whole density range.

The vertical line indicates the maximal (minimal) global density
obtained on the outer (inner) circle (which are respectively
1.09 and 1.06 ped/m, as indicated in table \ref{table:tab_nb}).
This means that data corresponding to the inner circle that
are on the left of this vertical line correspond to densities
that are far from average, i.e. they can be observed only
if there are strong fluctuations of the density.
When such a fluctuation toward low density occurs,
a pedestrian suddenly has a larger spatial headway.
Because he has a nonvanishing reaction time and
a finite acceleration, he will adapt his velocity
to this new situation with some delay.
It is also possible that the pedestrian has enough
information to anticipate that this fluctuation
of the spatial headway
will be short-lived and thus to decide not to adapt
to it.
In both cases, as a result, 
his velocity will be lower
than it would be in a stationary state.
This is what we observe: the values coming from
the inner circle data that are on the left of the vertical line 
correspond to lower velocity values.
This should not be an effect of the geometry
but rather the sign that the velocity has not fully
relaxed to its stationary value for these data.
Since these fluctuations in density are rather rare events, 
these points have a low weight if we take the average
over all data: the average coincides with the outer
circle values.

In a similar way, on the right of the vertical line,
the average over all data coincides with the inner circle
data. Large densities can be observed on the outer circle
only if there are density fluctuations.
Again, a pedestrian submitted to a fluctuation of
his spatial headway toward small distances will not
react immediately, and as a result the velocity
corresponding to these fluctuations will be higher
than in the stationary state.
This effect may here be enhanced by the fact that, on the outer
circle, the pedestrians can see quite far ahead.
The pedestrian can all the more
anticipate that his predecessor will accelerate
again and thus accept to come closer to him without
decreasing much his own velocity.
Indeed, the decrease in the radius that was observed
in figure \ref{fig:radius_markers} occurs precisely
in a density range from 1.2 to 2 ped/m, i.e. a density
range corresponding to transient fluctuations (as
these densities are above the global density).
This can be related, as mentioned earlier,
to a strategy to see further ahead and to
limit the inconvenience of coming very close to the
predecessor.
Such a strategy is not so interesting on the inner circle,
as an increase in the radius does not improve so much
the visual field, and does not shorten the trajectory.

Thus we can explain the branches going away from the
average behavior on figure \ref{fig:fd_markers_inout}
as being a signature of transient
behaviors, when pedestrians' velocities do not have
time to relax toward their stationary value.
Besides, the weight of such transient is low
compared to the remaining set of data.

The velocity-density points averaged on all data
(blue diamond symbols on figure \ref{fig:fd_markers_inout})
show no discontinuity when going from the outer
to the inner circle data.

The fundamental diagram averages over both stationary
and transient behaviors, which cannot be easily separated.
For low global densities, though fluctuations may arise
from individual variations in the walking speed, 
the distribution of pedestrians is more or less
homogeneous along the circle.
When the global density becomes quite high, stop-and-go waves
are produced.
The maximal global density on the inner circle is 1.86 ped/m.
Thus densities larger than this value are only obtained within
compression waves. However, jams are quite stable,
and even if there are some transient behaviors when
people enter or leave the jam, inside the jam we would
a priori
consider that pedestrians are in a stationary state,
i.e. that they have adjusted their velocity to the
space available in front of them.

We have plotted the obtained averaged fundamental diagram together
with the fundamental diagrams from the global and local measurements
in figure \ref{fig:fd_global2}. It is remarkable that even at densities as large as
$3$ ped/m, the velocity does not vanish but rather
reaches a limiting value.
This is consistent with other observations in two-dimensional
flows showing that, in contrast with car traffic, pedestrians
can move on even at very high densities \cite{helbing_j_a07}.

\section{Phase transitions in the following behavior}
\label{sect_transition}

\subsection{Experimental evidence of three dynamical regimes}
\label{sect_phases}

\begin{figure}[t] %  figure placement: here, top, bottom, or page
  \centering
      \includegraphics[width=0.48\textwidth]{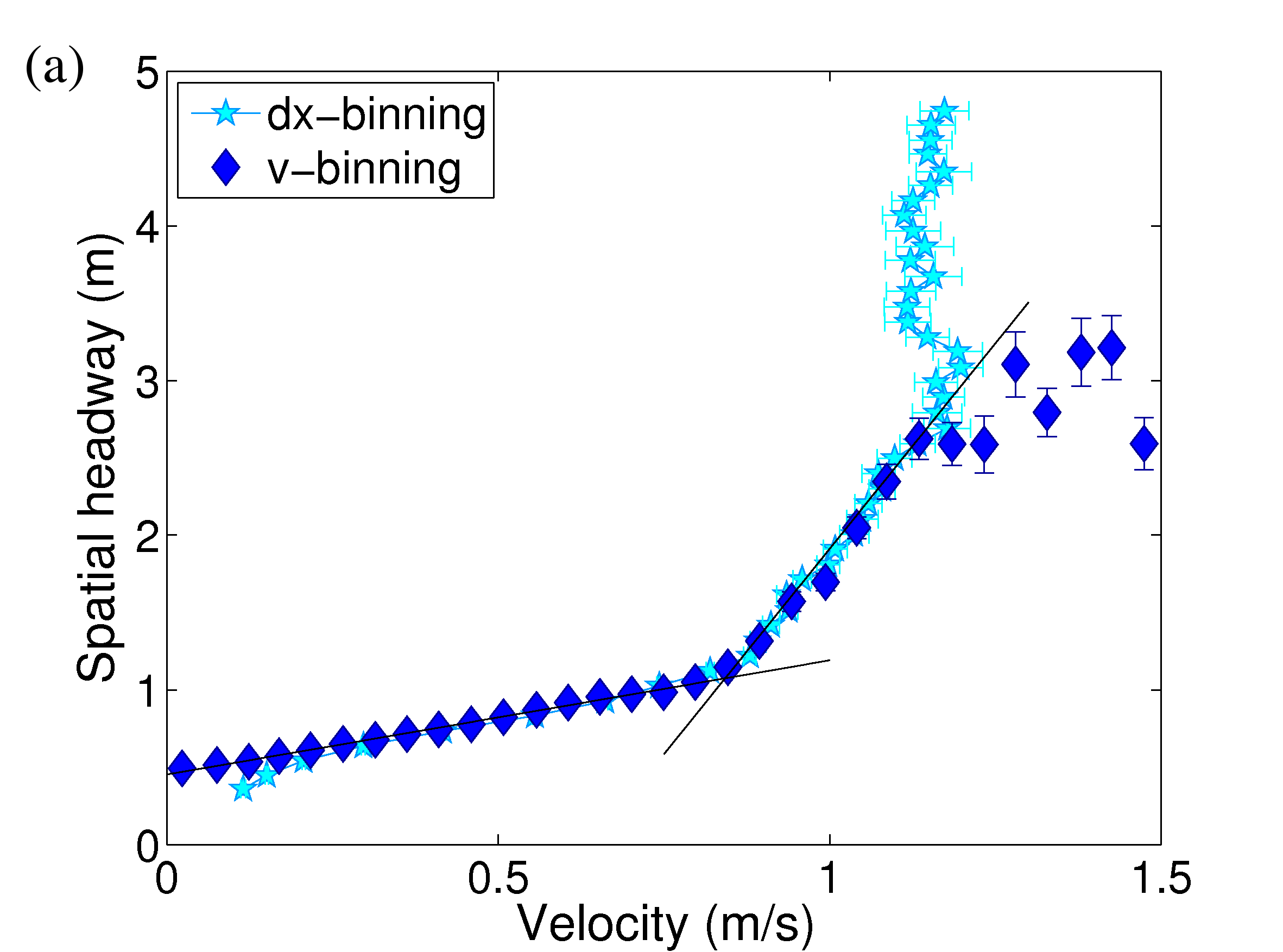}
\includegraphics[width=0.48\textwidth]{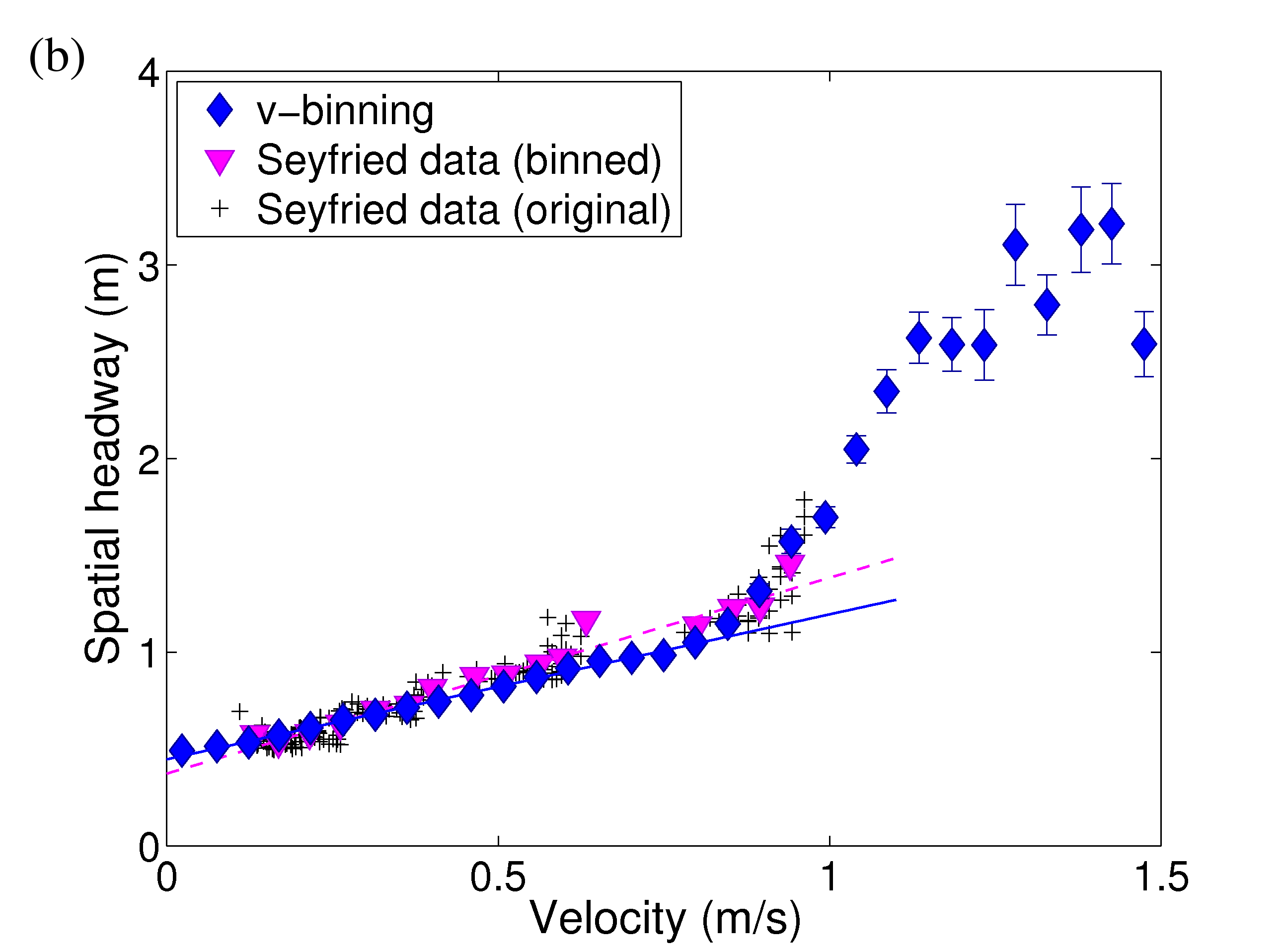}
   \caption{
   (Color online)
Spatial headway $\dx=1/\rho$ as a function of velocity $v$.
Each figure is obtained
as the result of a binning procedure over a large number of measurements.
Each measurement corresponds to an average over 0.5s, for one given pedestrian.
On (a), experimental data are plotted
using both a horizontal (cyan stars)
and a vertical (blue diamonds) binning procedure.
Indeed, a binning on spatial headway values is more
appropriate when the velocity saturates.
Solid lines are fits of the data in the strongly and
weakly constrained regimes.
On (b), 
our data (blue diamonds) are
compared with the measurements obtained
in \cite{seyfried05} (magenta triangles). 
The lines (respectively blue and red online) correspond to fits
of our data and Seyfried's data, both restricted to velocity $<$ 0.8 m/s.
 }
   \label{fig:culture}
\end{figure}

In \cite{seyfried05}, Seyfried {\em et al.} have observed that
the velocity of individuals is related to the spatial headway (distance
to the predecessor) by a linear relation.
With our experimental data, we are able to cover a larger 
range of velocities (or, equivalently, of instantaneous densities)
and we find that there are actually several linear regimes, 
with different slopes (figure
\ref{fig:culture}-a) \footnote{When a binning on velocity is used, it is not possible
to observe the saturation of the velocity at low densities.
The vertical part of the relation shown in figure \ref{fig:culture}-a
can be seen only with a binning on the
spatial headway, while for the almost horizontal 
part at low densities, a binning on the velocity is more appropriate.
Thus we superimpose the results obtained for both types of aggregating procedures.}.

Three regimes can clearly be distinguished:
\begin{enumerate}[(a)]
\item For spatial headways greater than $3$m,
pedestrians walk with their preferred velocity and do not
seem to interact much with other pedestrians.
Thus we call this state {\em free} regime.
\item For spatial headways between $1.1$ and $3$m,
the velocity depends only weakly on the interdistance,
and pedestrians do not slow down much.
We shall refer to this regime as {\em weakly constrained}.
\item For spatial headways below $1.1$m (or equivalently,
instantaneous densities greater than 0.9 ped/m), the adaptation
of the velocity to the spatial headway is much more
pronounced. 
We shall refer to this regime as the {\em strongly constrained}
regime.
\end{enumerate}
The crossover between weakly and strongly constrained regimes
occurs for a velocity around 0.8 m/s.

Surprisingly, the adaptation in the strongly constrained regime
follows the same linear law even when the velocity comes very close to zero.

When the velocity tends to zero, pedestrians remain at a certain
minimal distance $\dx_0$ from each other. It must be remembered that
the spatial headway is here defined between centers of mass, while pedestrians
have a certain thickness.

The slope in each linear regime has the dimension
of time. Following \cite{seyfried05}, we interpret it
as a sensitivity to the spatial headway.
It describes how strongly the pedestrian adapts his velocity
to the available space in front of him.
It can also be seen as an adaptation time, as it is the time
available to react before being at the 
minimal distance $\dx_0$ from the current position of the predecessor.

It is natural that the adaptation time
decreases when the spatial headway decreases, as
possible collisions are more imminent.
The surprise here comes from the fact that the characteristic adaptation time
does not vary smoothly but only takes three values with
sharp transitions between them.
Table \ref{tab_reaction_time} summarizes
the values of the adaptation time
in the three regimes.

\begin{table}[h]
\begin{tabular}{|c|c|}
\hline
\hspace{0.2cm} Regime type \hspace{0.2cm} & \hspace{0.2cm} adaptation time [s] \hspace{0.2cm} \\
\hline
Free & $13.7$ \\
\hline
Weakly constrained & $5.32$ \\
\hline
Strongly constrained & $0.74$ \\
\hline
\end{tabular}
\caption{Adaptation time (or sensitivity) in the three regimes.
Each adaptation time is obtained from a linear fit
of the spatial headway - velocity relation in figure
\ref{fig:culture} (a). For the free regime we fitted 
data with interdistance $\dx>3.2$ m. However, 
as the curve is almost vertical and statistical errors obviously
large, the numerical value is not meaningful, we can just conclude
that the velocity saturates 
for these spatial headways.}
\label{tab_reaction_time}
\end{table}

\subsection{Comparison of fundamental diagrams
from different cultures}
\label{sect_culture}

A comparison between data obtained in Germany and
India \cite{chattaraj_s_c09} showed that walking 
characteristics could be culturally dependent.
Before comparing our French data to those,
it should be noticed on figure \ref{fig:culture} (b)
that the beginning of the weakly constrained regime 
was already visible in the case of Seyfried's data \cite{seyfried05,seyfried07b},
and that, interestingly enough, the transition occurs around the
same velocity as for the French data.

When measuring the slope for the strongly constrained regime,
one should be careful not to include data from the weakly
constrained regime.
However, though this was not taken into account in \cite{seyfried05},
the difference with the new slope that we
have measured from their data
is small (see table \ref{tab_compare}).
There is still some uncertainty on the slope for the German data,
as there are not too many data for $v>0.6$m/s.
Besides, it seems that there was a large statistical
fluctuation around a velocity of $0.6$ m/s (see Seyfried's data
in figure \ref{fig:culture}-b) which may bias the result.

\begin{table}[h]
\begin{tabular}{|c|ccc|}
\hline
& \hspace{0.2cm}France \hspace{0.2cm} & 
\hspace{0.2cm}Germany \hspace{0.2cm}&
\hspace{0.2cm} India \hspace{0.2cm} \\
\hline
Intercept a[m] & 0.45 & 0.37 (0.36) & (0.22)\\
\hline
Slope b[s] & 0.75 & 1.01 (1.04) & (0.89)\\
\hline
\end{tabular}
\caption{Slope (adaptation time) and intercept with the zero velocity axis
(minimal distance $\dx_0$)
of a linear fit of the distance headway - velocity relation
in the strongly constrained regime.
The French and corrected German values were obtained from a linear fit
over not-binned (original) data
for velocities smaller than 0.8 m/s
in figure \ref{fig:culture} (b).
We indicate the values published in \cite{chattaraj_s_c09}
in parenthesis.}
\label{tab_compare}
\end{table}

Still, it seems that for a given spatial headway,
velocities are  systematically slightly higher in the French case
than in the German case.
Also, a higher intercept value $\dx_0$, which can be seen as a minimal personal 
space \cite{chattaraj_s_c09},  is observed in the French data.
A high intercept means that pedestrians do not accept too small
distances even in strong jamming.
Consistent with these two facts,
the French adaptation time is smaller than the German one,
which means that their reaction at small distances is stronger.

Finally, for Indian data, we were not able to check
whether the linear fit was performed
only on the strongly constrained data. From the available 
values for the slope and the intercept of the linear fit
performed in \cite{chattaraj_s_c09}, we can conclude 
that the minimal personal space takes the lowest value 
for Indian pedestrians, while their adaptation time to the 
change of spatial headway is in between 
the French and the German values.

To say it otherwise, in view of the current available data,
we could say that the strategy of Indian
pedestrians facing jamming is rather to accept small distances,
the strategy of French pedestrians is rather to break hard,
and the strategy of German pedestrians is to walk at lower velocities.

However, it should be underlined that these cultural
differences are weak, and that remarkably, in the
three countries considered, velocity varies linearly
with distance for distances below one meter.
There seems to be a kind of universality in pedestrian's
behavior, that would be interesting to probe on a larger
range of densities.

\section{Conclusion}
\label{sect_conclusion}

In this paper, we present some analysis of new experiments
for pedestrians walking on a line using high precision
motion capture.
We have focused on the fundamental diagram measurements
and shown that various definitions for the velocity and
the density could be used, bringing complementary informations.
In particular, our data allowed a measurement of the instantaneous
fundamental diagram.
We have shown that 
such a fundamental diagram actually
aggregates both stationary and non-stationary behavior.
Indeed, for a given experiment, the results obtained far from 
the global density, i.e.,\ for fluctuations, show quite a different behavior
from the average behavior,
as the velocity is for
a while not adapted to the available space - a feature
that can be seen as fluctuations around the mean behavior
in the fundamental diagram.

Our main result is to show that the velocity - spatial headway
relation allows to distinguish three linear regimes,
that we call free, weakly constrained, and strongly constrained.
The crossovers between these regimes are quite sharp.
The sensitivity, or adaptation time, characterizes how strongly
a pedestrian adapts his velocity to the space available in front
of him.
Instead of changing continuously, the adaptation time
takes only three discrete values, corresponding to the three regimes.
This observation should give new directions for further modeling.

Any comparison between different sets
of data should be careful at covering enough densities
to distinguish these various regimes.
Here we have compared our French data obtained in the 
strongly constrained regime with German \cite{seyfried05}
and Indian \cite{chattaraj_s_c09} data.
When facing congestion, a 
pedestrian can choose to slow down preventively, or to react more rapidly to changes,
or to accept smaller distances with his predecessor.
Cultural background can enhance one of these strategies,
though further careful experiments would be needed
to confirm the trends that we observe.
However, cultural effects are on the whole rather
weak and pedestrians' behavior appears mostly
universal at the small distances for which a comparison
could be made. Further studies covering a larger
density range would be of interest, in particular
to explore the robustness of the transitions
that we have found.

A thorough study of the stepping behavior of pedestrians,
depending on the density and velocity, will be published in 
a future paper \cite{jelic11b}.

In a separate paper, an analysis of the following behavior of pedestrians
has led to the development of a microscopic model, 
which is shown to reproduce quite well the stop-and-go waves
observed in the experiment \cite{lemercier12a,hua12}.
Besides, its extension to a quasi-1D multi-lane model to simulate
unidirectional flow in a corridor turns out to give
results as realistic as alternative methods \cite{lemercier12a}.

\begin{acknowledgments}

This work has been supported by the French ``Agence Nationale pour la Recherche (ANR)'' in the frame of the
Contract ``{\sc PEDIGREE}'' (Contract No. ANR-08-SYSC-015-01).
The {\sc PEDIGREE} Project is financed by the French ANR and
involves four research teams in Rennes (INRIA), Toulouse
(IMT, CRCA), and Orsay (LPT).
Experiments were organized and realized by the {\sc PEDIGREE} 
partnership \cite{pedigree_info} at University Rennes 1, with the help of
the laboratory M2S from Rennes 2. We are in particular 
grateful to Armel Cr\'etual, Richard Kulpa, Antoine Marin, and 
Anne-H\'el\`ene Olivier for their help during the experiments. 

A.J. acknowledges support from the RTRA Triangle de la 
physique (Project 2011-033T).

\end{acknowledgments}

\bibliographystyle{unsrt}

\end{document}